\documentclass[twocolumn,pra,floatfix,citeautoscript,nofootinbib,superscriptaddress]{revtex4}
\usepackage{amsbsy}
\usepackage{latexsym,epsfig,graphicx}
\usepackage{dcolumn}
\usepackage{graphicx}
\usepackage{subfigure}
\usepackage{comment}
\usepackage{color}
\usepackage{bm}
\usepackage{mathrsfs}
\usepackage{amsfonts}
\usepackage{amsmath}
\usepackage{color}
\usepackage{amssymb}
\usepackage{xspace}
\usepackage{epstopdf}
\usepackage{tabularx}
\usepackage{longtable}
\usepackage[colorlinks=true, letterpaper=true, pdfstartview=FitV, linkcolor=blue, citecolor=blue, urlcolor=blue]{hyperref}
\usepackage[normalem]{ulem}

\allowdisplaybreaks[4]
\pdfoutput=1

\begin{document}

\title{Two-dimensional non-Hermitian topological phases induced by
asymmetric hopping in a one-dimensional superlattice}
\author{Junpeng Hou}
\affiliation{Department of Physics, The University of Texas at Dallas, Richardson,
Texas75080-3021, USA}
\author{Ya-Jie Wu}
\affiliation{Department of Physics, The University of Texas at Dallas, Richardson,
Texas75080-3021, USA}
\affiliation{School of Science, Xi'an Technological University, Xi'an 710032, China}
\author{Chuanwei Zhang}
\email{chuanwei.zhang@utdallas.edu}
\affiliation{Department of Physics, The University of Texas at Dallas, Richardson,
Texas75080-3021, USA}

\begin{abstract}
Non-Hermitian systems can host topological states with novel topological
invariants and bulk-edge correspondences that are distinct from conventional
Hermitian systems. Here we show that two unique classes of non-Hermitian 2D
topological phases, a 2$\mathbb{Z}$ non-Hermitian Chern insulator and a $%
\mathbb{Z}_{2}$ topological semimetal, can be realized by tuning staggered
asymmetric hopping strengths in a 1D superlattice. These non-Hermitian
topological phases support real edge modes due to robust $\mathcal{PT}$%
-symmetric-like spectra and can coexist in certain parameter regime. The
proposed phases can be experimentally realized in photonic or atomic systems
and may open an avenue for exploring novel classes of non-Hermitian
topological phases with 1D superlattices.
\end{abstract}

\maketitle

\section{Introduction}

In the past few decades, topological states of matter have been extensively
studied in various physical systems because of their unusual properties and
significant applications in quantum devices and information processing \cite%
{XiaoD2010,ChiuCK2016}. While the study has mainly focused on solid-state
materials \cite{WangJ2017}, ultra-cold atomic gases provide another platform
for realizing topological states with high tunability and controllability 
\cite%
{AidelsburgerM2013,AtalaM2013,AidelsburgerM2014,JotzuG2014,LohseM2015,NakajimaS2016,GoldmanN2016,CooperNR2019}%
. Furthermore, the concept of topology has been extended to classic systems
govern by wave equations such as photonics \cite%
{HaldaneFDM2008,WangZ2009,LuL2010,LuL2014,KhanikaevAB2017,SirokiG2017,BahariB2017}%
, acoustics \cite{HeC2016,HeH2018}, and electric circuits \cite{ImhofS2018},
yielding many interesting topological states.

One significant feature of ultra-cold atomic gases and these classical
systems comparing to quantum materials is their capability of controllably
inducing non-Hermiticity (e.g., gain and loss), which makes them excellent
platforms for exploring novel non-Hermitian physics, such as unidirectional
transportation \cite{WuJH2014}, spontaneous $\mathcal{PT}$ symmetry breaking
with exceptional points \cite{KreibichM2014,LiJ2019}, fast eigenstate
transition \cite{MostafaviF2019} and novel superfluidity \cite%
{PanL2019,YamamotoK2019}, etc. In particular, significant effects and
applications have been proposed or demonstrated in non-Hermitian photonics 
\cite%
{MakrisKG2008,RegensburgerA2012,PengB2014,ZeunerJM2015,FengL2017,HouJ2019}.

The combination of topology and non-Hermiticity leads to the emergence of
novel topological effects, such as anomalous edge states, non-Bloch waves,
non-Hermitian skin effects, etc \cite%
{LeeTE2016,KunstFK2018,YaoS2018N,YaoS2018,LieuS2018,ShenH2018,KawabataK2019,JinL2019,LinS2019}%
. In experiments, photonic \cite%
{MalzardS2015,StJeanP2017,BandresMA2018,PartoM2018,TakataK2018,ZhouH2018,CerjanA2018,HouJ2018,LuoXW2019}
and atomic \cite{XuY2017,LiuT2019} systems are leading platforms for
realizing non-Hermitian topological states. Although in most work the
non-trivial topology is attributed to the Hermitian part of the Hamiltonian,
it has been recently proposed that topological states may be solely induced
by non-Hermiticity \cite{TakataK2018} with a focus on higher-order
topological insulators \cite{LiuT2019,LuoXW2019,LeeChH2018}. The
classification of non-Hermitian topological phases has been proposed through
a reduction from AZ classes \cite{GongZ2018}, which remains incomplete
because a random non-Hermitian matrix belongs to a broader BL class \cite%
{BernardD2001}. More recently, a comprehensive classification of
non-Hermitian systems has been achieved through introducing the complex AZ$%
{}^{\dagger}$ class \cite{KawabataKSymmetry2019}.

In Hermitian systems, it is known that 2D topological phases, such as a
Chern insulator, can be simulated using a 1D lattice with staggered hopping
or on-site potential \cite{LangLJ2012} characterized by an additional
periodic parameter. The experimental realization of such 2D topological
phases in momentum-parameter space could be significantly simpler, comparing
to their 2D lattice counterparts. In this context, two natural questions
arise for non-Hermitian systems: i) Can 2D non-Hermitian topological phases
be realized using 1D lattices with solely non-Hermitian effects? 2) If so,
is there any unique class of non-Hermitian topological phases that are
difficult to realize in Hermitian systems?

In this work, we address these two important questions by showing that two
unique symmetry-protected non-Hermitian topological phases can be realized
in a 1D superlattice with staggered asymmetric hopping. Our main results are:

\textit{i}) The 1D superlattice hosts a 2D topological insulator phase
characterized by a $2\mathbb{Z}$ Chern number, which is protected by certain
symmetry that is unique to non-Hermitian systems \cite{BernardD2001}. Such a
phase is difficult to access in Hermitian systems because it is absent in
the AZ classification and requires a crystalline symmetry as well as a
global $\mathbb{Z}_{2}$ symmetry \cite{ShiozakiK2014}.

\textit{ii}) A topological semimetal phase hosting complex Dirac points and
zero-energy modes is found to be associated with a $\mathbb{Z}_{2}$
invariant, which is extracted from the normalized Berry phase for
non-Hermitian systems.

\textit{iii}) The system may support edge states with real energy due to a
robust $\mathcal{PT}$-symmetric-like phase. In most non-Hermitian systems
with on-site gain and loss, the $\mathcal{PT}$-symmetric phase becomes
fragile with increasing system size. In contrast, the $\mathcal{PT}$%
-symmetric-like phase here is robust to varying chain length. Furthermore,
these two topological phases can coexist in proper parameter spaces.

\section{2D non-Hermitian topological phases in 1D superlattice with
asymmetric hopping}

\subsection{Model Hamiltonian}

We consider a lattice model with nearest neighbor hopping, which can be
described by a tight-binding Hamiltonian $H_{r}=\sum_{i}\left( t_{i,i+1}\hat{%
c}_{i}^{\dagger }\hat{c}_{i+1}+t_{i+1,i}\hat{c}_{i+1}^{\dagger }\hat{c}%
_{i}\right) +V_{i}\hat{c}_{i}^{\dagger }\hat{c}_{i}$. Here $\hat{c}%
_{i}^{\dagger }$($\hat{c}_{i}$) is the creation (annihilation) operator of
local modes at site $i$, and the hopping term is non-Hermitian $%
t_{i,i+1}\neq t_{i+1,i}^{\ast }$. For simplicity of the presentation, we
assume uniform on-site potential and staggered hopping terms $%
t_{i,i+1}=1+\lambda \cos (2\pi \alpha i+\phi _{L})$ and $t_{i+1,i}=1-\lambda
\cos (2\pi \alpha i+\phi _{L})$, where $\lambda ,\alpha \in \mathbb{R}$. We
consider rational $\alpha $ that can be written as the quotient of two
relatively prime integers $\alpha =p/q$, $p,q\in \mathbb{Z}$. Without loss
of generality, $p,q$ are assumed to be positive and $p< \lceil q/2\rceil$.
The corresponding Bloch Hamiltonian $H(k,\phi _{L})$ for the superlattice is
a $q\times q$ matrix in the Brillouin zone $|k|\leq \pi /q$ with non-zero
entries $H_{j,j}=V_{j}$, $H_{j,j+1}=t_{j,j+1}$, $H_{j+1,j}=t_{j+1,j}$ for $%
j+1<q$, $H_{1,q}=t_{1,q}e^{-iqk}$ and $H_{q,1}=t_{q,1}e^{iqk}$.

The resulting Bloch Hamiltonian $H(k,\phi _{L})$ preserves time-reversal
symmetry $\mathcal{T}_{k}H(k,\phi _{L})\mathcal{T}_{k}^{-1}=H^{\ast }(k,\phi
_{L})=H(-k,\phi _{L})$ with $\mathcal{T}_{k}=K$ and $K$ the complex
conjugate, therefore the real (complex) part of the band is symmetric
(antisymmetric) about $k=0$. Particle-hole symmetry yields $\mathcal{P}%
_{\phi _{L}}H(k,\phi _{L})\mathcal{P}_{\phi _{L}}^{-1}=H^{T}(k,-\phi _{L})$,
where $\mathcal{P}_{\phi _{L}}$ has a permutation representation $%
(q,1)(q-1,2)(q-2,3)...$ in 2-cycle forms.

\begin{figure}[t]
\centering
\includegraphics[width=0.48\textwidth]{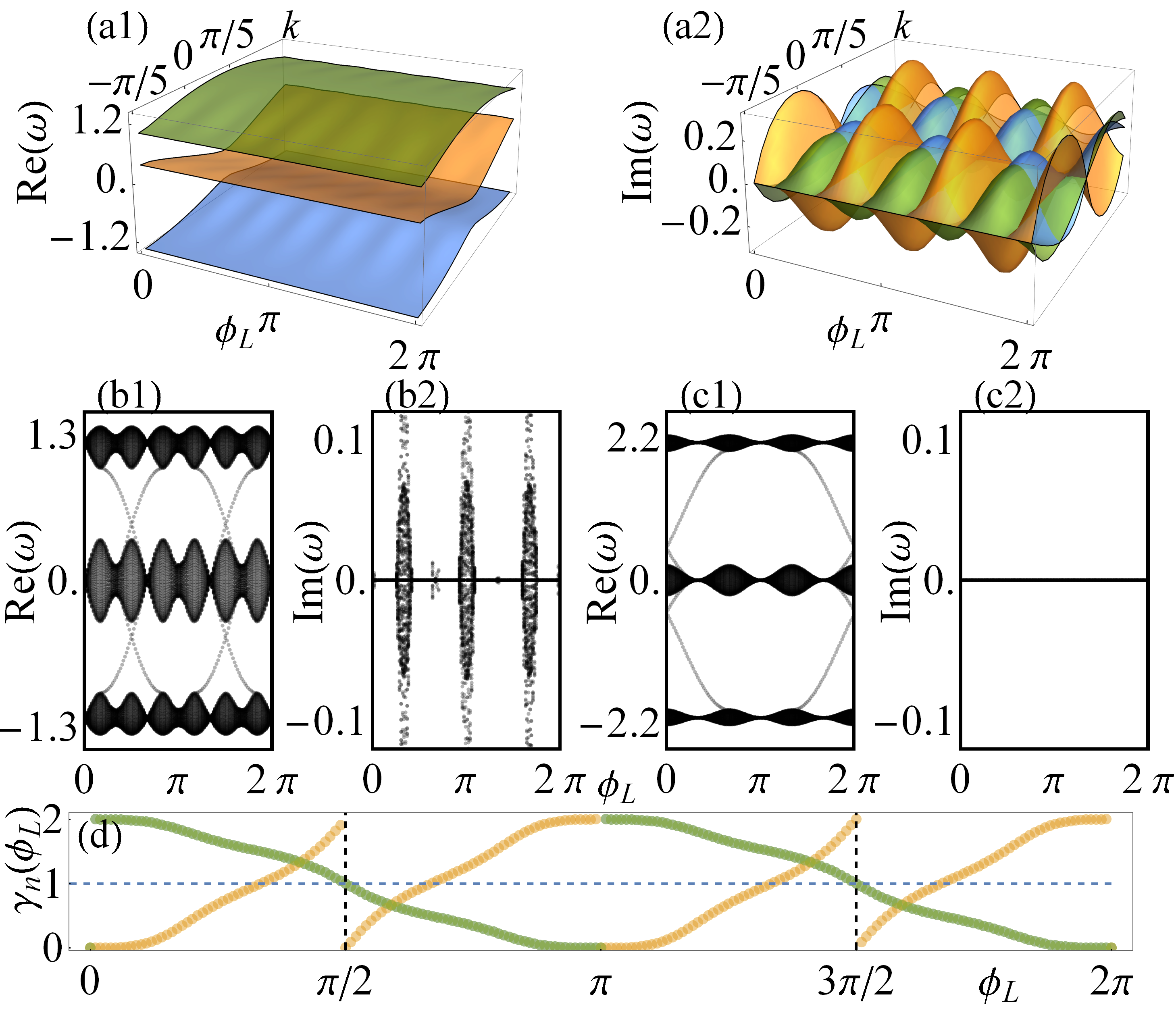}
\caption{$2\mathbb{Z}$ Chern insulator phase for $\protect\alpha =1/3$. (a)
2D band structure in momentum-parameter space when $\protect\lambda =1$. The
Chern numbers for each band are $-2$, $4$ and $-2$ (from bottom to top). (b)
Open-boundary spectra with respect to $\protect\phi _{L}$ for $\protect%
\lambda =1$ and $N_{L}=120$. (c) Similar to (b) but plotted with $\protect%
\delta \protect\phi _{L}=\protect\pi $, which leads to a $\mathbb{Z}$ Chern
insulator with odd Chern number. (d) Normalized Berry phase $\protect\gamma %
_{n}(\protect\phi _{L})$ for bulk bands in (b). The dashed vertical lines
indicate inversion-symmetric points.}
\label{FigChernIn}
\end{figure}

\subsection{\emph{$2\mathbb{Z}$} Chern insulator}

We start with the simplest case $\alpha =1/3$, which represents a Chern
insulator, as shown in Fig.~\ref{FigChernIn}(a). The Bloch band is gapped by
its real part in the entire momentum space for any $\phi _{L}$ and exhibits
the symmetry $\omega (k)=\omega (-k)^{\ast }$. We also plot the
open-boundary spectrum with varying $\phi _{L}$ in Fig.~\ref{FigChernIn}(b).
Since the open-boundary Hamiltonian $H_{O}(\phi _{L})=H_{O}(\phi _{L}+\pi
)^{T}$, the spectrum is symmetric about $\phi _{L}=\pi $. When the lattice
size satisfies $N_{L}/q\in \mathbb{Z}$, the Hamiltonian on a cylindrical
geometry enjoys a chiral symmetry $\mathcal{C}_{O}H_{O}(\phi _{L})\mathcal{C}%
_{O}^{-1}=H_{O}(\phi _{L})$ with $\mathcal{C}_{O}=I_{N_{L}/q}\otimes 
\mathcal{C}$, where $I_{n}$ is a $n\times n$ identity matrix and $\sigma
_{i} $ are Pauli matrices. Thus the open-boundary spectrum is symmetric
about $\omega =0$, as shown in Figs.~\ref{FigChernIn}(b) and (c). However,
the spectra shown in Figs.~\ref{FigChernIn}(b) and (c) may not obey these
symmetries exactly due to numeric errors, which are significantly enhanced
when diagonalize a non-Hermitian matrix.

Two pairs of \textquotedblleft chiral\textquotedblright\ surface waves
emerge in each gap [Fig.~\ref{FigChernIn}(b)] for varying $\phi _{L}$. When $%
\alpha =1/5$, similar results with two or four pairs of edge modes in any of
the four gaps are observed. In general, the number of the pair of edge
states is always even, as expected for a $2\mathbb{Z}$ Chern insulator
phase. The Hamiltonian satisfies $H(k,\phi _{L})=H^{T}(-k,\phi _{L}+\pi )$,
which leads to $H(k,\phi _{L})=H^{\dagger }(k,\phi _{L}+\pi )$ when combined
with time-reversal symmetry, yielding the $2\mathbb{Z}$ Chern insulator
phase. When an extra term $\delta \phi _{L}\in \mathbb{R}$ is added in the
asymmetric hopping $t_{i+1,i}=1-\lambda \cos (2\pi \alpha i+\phi _{L}+\delta
\phi _{L})$ that breaks this symmetry, only one pair of surface waves are
observed in each gap, as shown in Fig.~\ref{FigChernIn}(c).

The above 1D physics mimic the 2D integer quantum Hall effect in the $\left(
k,\phi _{L}\right) $ momentum-parameter space. The band topology can be
characterized by the Chern number 
\begin{equation}
C^{ab}=\frac{1}{2\pi }\int dkd\phi _{L}(\partial _{k}\mathcal{A}_{\phi
_{L}}^{ab}-\partial _{\phi _{L}}\mathcal{A}_{k}^{ab}),
\end{equation}%
where $a,b=L,R$, the Berry connections are defined as $\mathcal{A}%
_{k}^{ab}=-i_{a}\langle \psi (k,\phi _{L})|\partial _{k}\psi (k,\phi
_{L})\rangle _{b}$ and $\mathcal{A}_{\phi _{L}}^{ab}=-i_{a}\langle \psi
(k,\phi _{L})|\partial _{\phi _{L}}\psi (k,\phi _{L})\rangle _{b}$, and the
right and left eigenstates are defined as $H(k,\phi _{L})|\psi (k,\phi
_{L})\rangle _{R}=\omega (k,\phi _{L})|\psi (k,\phi _{L})\rangle _{R}$ and $%
H(k,\phi _{L})^{\dagger }|\psi (k,\phi _{L})\rangle _{L}=\omega ^{\ast
}(k,\phi _{L})|\psi (k,\phi _{L})\rangle _{L}$. Here the four Chern numbers $%
C^{ab}$ are equivalent \cite{ShenH2018} and $C^{RL}$ is numerically found to
be $-2$, $4$ and $-2$ from bottom to top bands in Fig.~\ref{FigChernIn}(a).
Thus the bulk topological invariant agrees well with the edge states through
usual bulk-edge correspondence. The Chern numbers are also consistent with
the 2$\mathbb{Z}$ Chern insulator and the detailed proof of exact 2$\mathbb{Z%
}$ quantization of Chern number can be found in Appendix \ref{AppA}.

The edge states are also closely related to the 1D topology. When combine
the particle-hole symmetry with $H(k,\phi _{L})=H^{T}(-k,\phi _{L}+\pi )$,
we obtain an inversion (glide) symmetry along $k$ ($\phi _{L}$), $\mathcal{I}%
_{k}H(k,\phi _{L})\mathcal{I}_{k}^{-1}=H(-k,-\phi _{L}+\pi )$. Such a
symmetry leads to high-symmetry points $\phi _{L}=\frac{\pi }{2}$ and $\frac{%
3}{2}\pi $, at which the edge states are degenerate and the corresponding 1D
normalized Berry phase $\gamma _{n}(\phi _{L})=\frac{1}{\pi }\oint dk%
\mathcal{A}_{k}^{n}$ is quantized. Here $\mathcal{A}_{k}^{n}=\frac{1}{2}%
\left( \mathcal{A}_{k}^{RL}+\mathcal{A}_{k}^{LR}\right) $ and $n$ represents
band index (see Appendix \ref{AppA} for details). We plot the Berry phase at
different $\phi _{L}$ in Fig.~\ref{FigChernIn}(d) and the color for each
band is the same as in Fig.~\ref{FigChernIn}(a). The top (green) and bottom
(blue) bands have the same normalized Berry phases, so only the green one is
visible.

We note that the normalized Berry phase is also quantized at $\phi _{L}=0$
and $\phi _{L}=\pi $, which is not a coincidence, but a result of a symmetry 
$\mathcal{Q}H(k,\phi _{L})\mathcal{Q}^{-1}=H^{\dagger }(-k,-\phi _{L})$ with 
$\mathcal{Q}=\mathcal{P}_{\phi _{L}}\mathcal{T}_{k}$ (see Appendix~\ref{AppA}
for details). A close observation reveals that $\gamma _{n}(\phi _{L})$
shows a period $\pi $ and is anti-symmetric to $\phi _{L}=\pi $. Note that
Fig.~\ref{FigChernIn}(d) also demonstrates a charge pumping process with
respect to $\phi _{L}$, where the accumulation of Berry phases from $\phi
_{L}=0$ to $2\pi $ gives the Chern numbers of three bands. This is
consistent with the $2\mathbb{Z}$ Chern number.

Finally, we remark that the summation of the normalized Berry phase always
vanishes (for both odd and even $q$), which suggests that the global Berry
phase used in studying similar 1D non-Hermitian systems \cite%
{LiangSD2012,TakataK2018} does not apply here. The topological invariant
(Chern number and the following $\mathbb{Z}_{2}$ invariant) always sums to
zero as the non-Hermiticity here does not affect the additivity in \textit{%
sum rule} \cite{HatsugaiY2004}. This is a well-defined 2D non-Hermitian
topological insulator characterized by a 2$\mathbb{Z}$ Chern number and
could be realized in a 1D lattice with only non-Hermitian modulations.

\begin{figure}[t]
\centering
\includegraphics[width=0.48\textwidth]{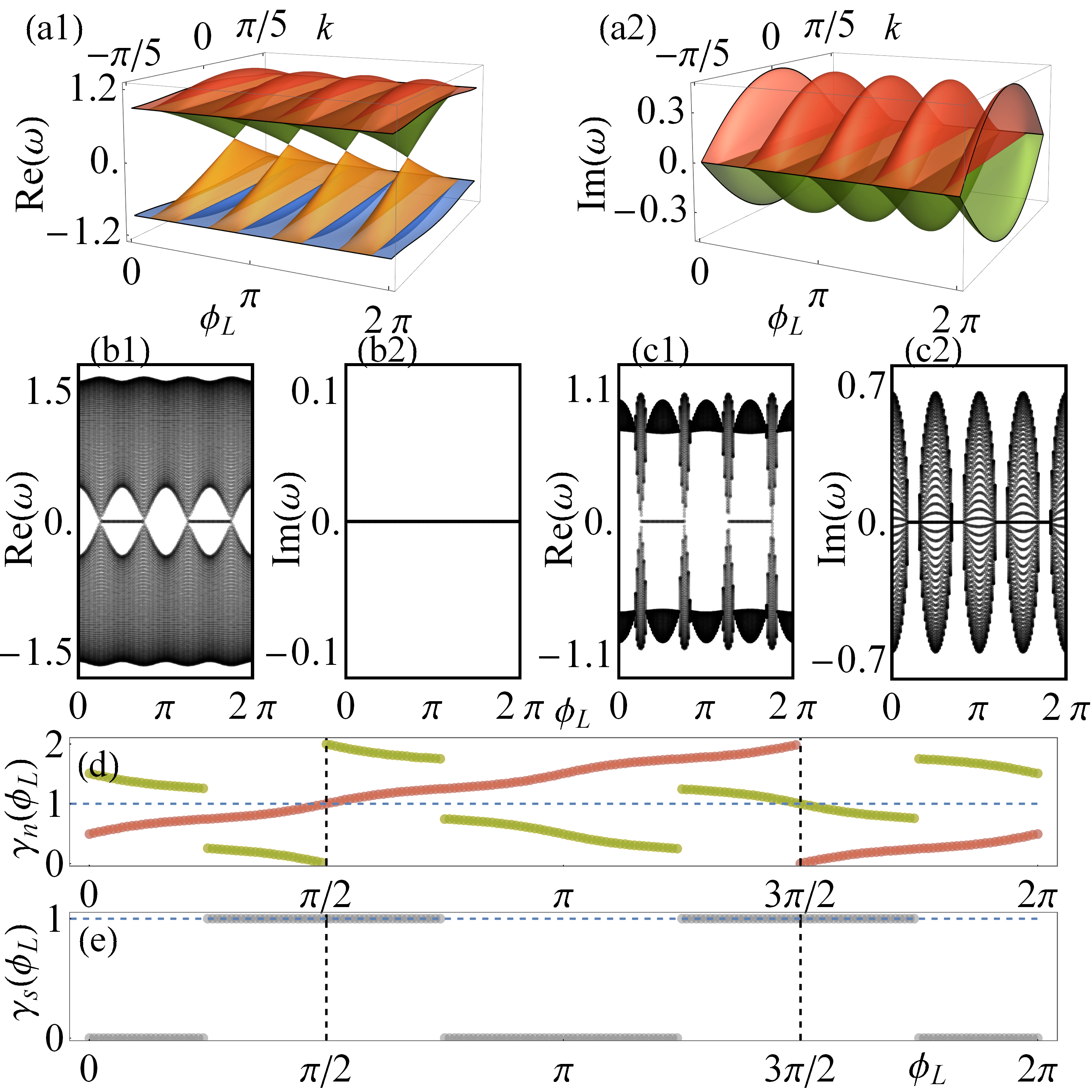}
\caption{$\mathbb{Z}_{2}$ topological semimetal phase when $\protect\alpha %
=1/4$. (a) 2D band structure in momentum-parameter space when $\protect%
\lambda =1.1$. Open-boundary spectrum with respect to $\protect\phi _{L}$
for (b) weak $\protect\lambda =0.8$ and (c) strong $\protect\lambda =1.2$
non-Hermitian modulations. We set $N_{L}=120$ in the spatial direction. (d)
Normalized Berry phase $\protect\gamma _{n}(\protect\phi _{L})$ for bulk
bands in (b). (e) The $\mathbb{Z}_{2}$ invariant computed from panel (d).
The dashed vertical lines indicate inversion-symmetric points.}
\label{FigTopoSemi}
\end{figure}

\subsection{$\mathbb{Z}_{2}$ topological semimetal}

We now consider $\alpha =1/4$ and the corresponding 2D band structure in
momentum-parameter space is shown in Fig.~\ref{FigTopoSemi}(a). The spectra
obeys the symmetry $\omega (k)=\omega (-k)^{\ast }$ and is symmetric about $%
\omega =0$. The latter results from an additional chiral symmetry $\mathcal{C%
}=I_{q/2}\otimes \sigma _{z}$ with $\mathcal{C}^{-1}H(k,\phi _{L})\mathcal{C}%
=-H(k,\phi _{L})$ when $q$ is even. Such a symmetry allows a finite band gap
at zero energy and is well-known for supporting zero-energy modes in 1D
Hermitian systems. The spectra is gapped everywhere in the
momentum-parameter space except at a few high-symmetry points, at which
emerge Dirac fermions. This naturally makes the system a semimetal. These
complex Dirac points are nontrivial and can be characterized by the
normalized Berry phase $\frac{1}{\pi }\oint_{L}dkd\phi _{L}\mathcal{A}%
_{k,\phi _{L}}^{n}=1$, where $L$ is a loop encircling the complex Dirac
point.

The open-boundary energy levels with respect to $\phi _{L}$ are given in
Figs.~\ref{FigTopoSemi}(b) and (c). The spectrum is real for a weak
modulation strength $\lambda$ and the two-fold degenerate edge modes have
exact zero and purely real energy despite the underlying non-Hermiticity of
the system. The gap closing points at certain inversion-symmetric points of $%
\phi _{L}$ and $k=0$ are the complex Dirac point in Fig.~\ref{FigTopoSemi}%
(a) and represents topological phase transition in the 1D picture.

The topological invariant associated with these zero modes comes from the
normalized Berry phases $\gamma _{n}(\phi _{L})$ that are computed
numerically in Fig.~\ref{FigTopoSemi}(d). Only the quantization of Berry
phase at inversion-symmetric points survives because it does not require a
gapped bulk (see Appendix \ref{AppA} for details). The two bands $\Re
(\omega )>0$ (the particle branch) have the same Berry phases as the two
with $\Re (\omega )<0$ (the hole branch). We define the $\mathbb{Z}_{2}$
invariant $\gamma _{s}(\phi _{L})=\sum_{j}\gamma _{n,j}(\phi _{L})\mod 2$,
where the summation runs over the hole branch and $\gamma _{s}(\phi _{L})=1/0
$ corresponds to topological/trivial phase. We find $\gamma _{s}(\phi _{L})=1
$ in the region $\frac{1}{4}\pi <\phi _{L}<\frac{3}{4}\pi $ and $\frac{5}{4}%
\pi <\phi _{L}<\frac{7}{4}\pi $, and 0 otherwise [see Fig.~\ref{FigTopoSemi}%
(e)], which agree with the edge states and gap closing in Figs.~\ref%
{FigTopoSemi}(a-c). Since the bulk bands could be degenerate at
high-symmetric points, we apply an infinitesimal $\delta \phi _{L}$ in the
hopping term $t_{i+1,i}$ to break the degeneracy so that both the Berry
phases and $\mathbb{Z}_{2}$ invariant can be well defined (see Appendix \ref%
{AppB}). Such a small perturbation does not break the chiral symmetry
therefore it does not affect the topological properties discussed in this
section.

Finally, we remark that an alternative definition of the $\mathbb{Z}_{2}$
invariant is to use the normalized non-Abelian Berry phase and take partial
trace over the hole branch. This does not require the gap-lifting term but
leads to the same result in Fig.~\ref{FigTopoSemi}(e).

\begin{figure}[t]
\centering
\includegraphics[width=0.48\textwidth]{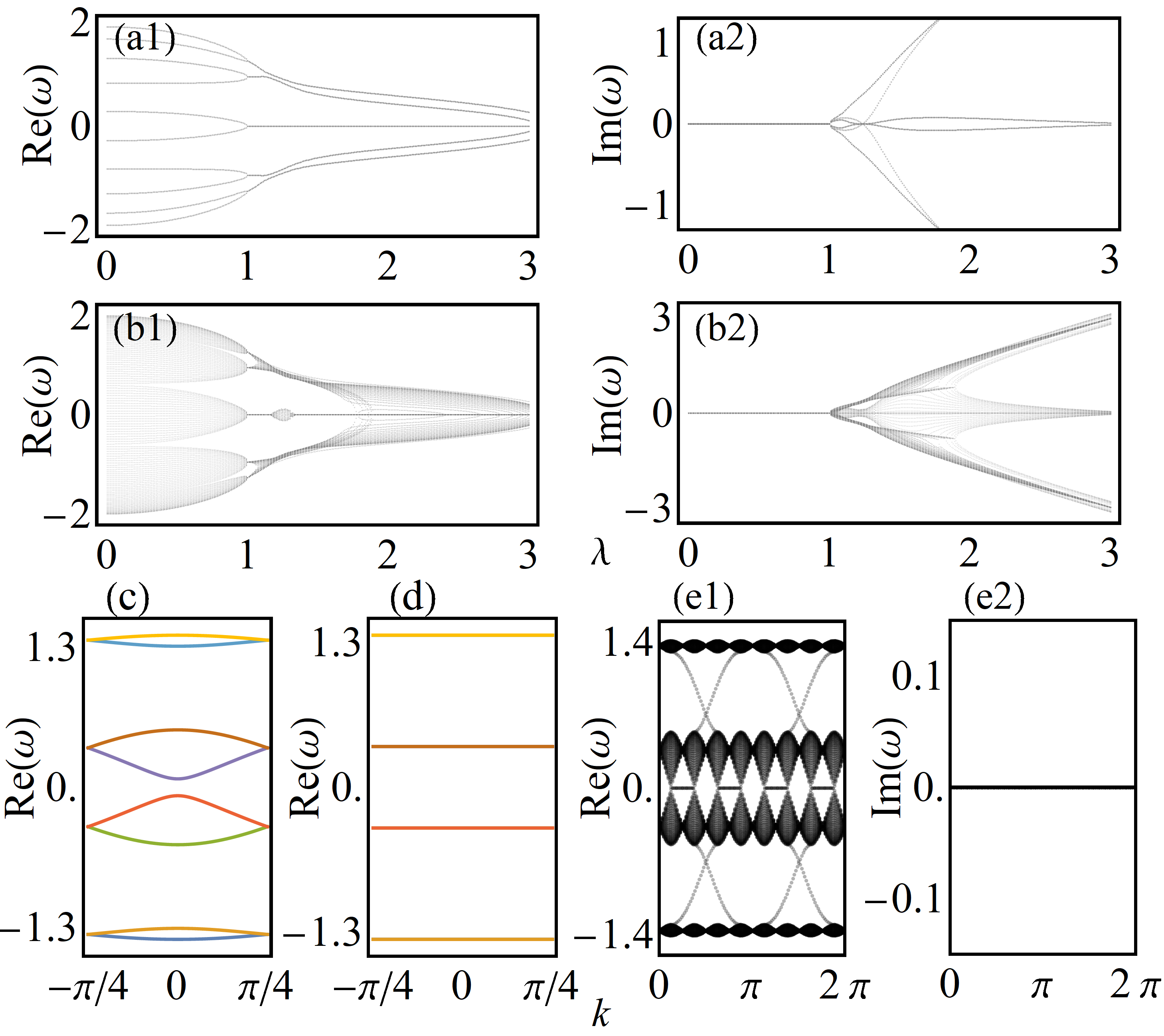}
\caption{(a) Open-boundary spectrum with respect to $\protect\lambda $ for $%
\protect\alpha =1/5$, $\protect\phi _{L}=0$ and $N_{L}=10$. (b) Similar to
(a) except $N_{L}=120$. (c) Bulk bands when $\protect\alpha =1/8$, $\protect%
\lambda =1$ and $\protect\phi _{L}=0.1\protect\pi $. (d) Flat bands at
high-symmetry points $\protect\phi _{L}=\protect\pi /4$. Other parameters
are the same as panel (c). (e) Coexistence of chiral surface wave and zero
modes at $\protect\alpha =1/8$ and $\protect\lambda =1$.}
\label{FigCoExist}
\end{figure}

\subsection{Robust $\mathcal{PT}$-symmetric-like and mixing phases}

The open-boundary spectrum is surprisingly real when the Bloch bands in the
Brillouin zone are imaginary, which is very different from most
non-Hermitian systems and is a result of the robust $\mathcal{PT}$%
-symmetric-like phase. The $\mathcal{PT}$ symmetry is generally fragile in
the sense that the critical value of driving term becomes very small when
the system size is very large \cite{YuceC2014} (see also Appendix \ref{AppC}%
). In contrast, the $\mathcal{PT}$-symmetric-like phase here is robust and
cannot be spontaneously broken when $\lambda <1$ regardless of the chain
length $N_{L}$. As an example, we show the spectrum of an open chain with
different length $N_{L}$ in Fig.~\ref{FigCoExist}(a) and (b). The spectrum
becomes complex at a fixed modulation strength $\lambda =1$ for both $%
N_{L}=10$ and $N_{L}=120$. This notable feature of robust $PT$%
-symmetric-like phases ensures the reality of both bulk spectrum and edge
modes in an extended parameter space.

To get insights of the $\mathcal{PT}$-symmetric-like phase, we consider
two-level systems with the following two non-Hermitian Hamiltonians $H_{%
\text{gl}}=\sigma _{y}+i\lambda _{\gamma }\sigma _{z},\lambda _{\gamma }\in 
\mathbb{R}$ and $H_{\text{ah}}=\sigma _{x}+i\lambda _{\mu }\sigma
_{y},\lambda _{\mu }\in \mathbb{R}$. The former one defines the simplest $%
\mathcal{PT}$-symmetric model with balanced gain and loss while the latter
one corresponds to the asymmetric hopping discussed in this work. It is
obvious that $H_{\text{ah}}$ is equivalent to $H_{\text{gl}}$ through basis
rotation $\sigma _{x}\rightarrow \sigma _{z}$, $\sigma _{y}\rightarrow
\sigma _{x}$ and $\sigma _{z}\rightarrow \sigma _{y}$, which explains the $%
\mathcal{PT}$-symmetric-like phase in our system.

The $2\mathbb{Z}$ Chern insulator exists for any $q>2$, while the
topological semimetal requires chiral symmetry and even $q$, which suggest
that these two distinct phases can mix for a large even $q$. We consider $%
\alpha =1/8$ and the Bloch band is plotted in Fig.~\ref{FigCoExist}(c),
which also exhibits flat band at high-symmetry points in the parameter space
as shown in Fig.~\ref{FigCoExist}(d). The real open-boundary spectrum for
the mixed phase is shown in Fig.~\ref{FigCoExist}(e). The topological
properties can be similarly characterized by taking into account of band
degeneracy (see Appendix \ref{AppD} for details).

\section{Discussion and conclusion}

Our model uses non-Hermitian nearest-neighbor hoppings on a lattice
structure with uniform on-site potential. This could be realized in
platforms like cold atoms in optical lattice or array of coupled waveguides.
Many schemes have been proposed to realize asymmetric hoppings in both
platforms \cite{LonghiS2015,LiuT2019}. A few sites with small period $q=2$
or $4$ is sufficient for versatile topological phases (see Appendix~\ref%
{AppB}) and it is readily accessible in current experiments. The chiral edge
waves and zero-energy modes have real energies in broad parameter spaces,
therefore they are free of dissipations/amplifications over time and are
easier to observe in experiments comparing to imaginary edge modes in many
other models \cite{YuceC2014,TakataK2018}.

There are many other non-Hermitian topological phases that could be explored
in this system by considering, for instance, other types of non-Hermitian
hoppings, staggered $V_{i}$, tilted lattice, and long-range hopping. These
new ingredients could bring richer physics as demonstrated already in
certain bipartite superlattice \cite{LieuS2018,LeeTE2016,YaoS2018}.

In conclusion, we show that two unique non-Hermitian 2D topological phases
-- a 2$\mathbb{Z}$ Chern insulator and a $\mathbb{Z}_{2}$ topological
semimetal -- can be realized in a 1D lattice with staggered non-Hermitian
hoppings. These phases can be experimentally realized in photonic or atomic
systems and may open an avenue for exploring novel classes of non-Hermitian
topological phases with 1D superlattices.

\textbf{Note added.} During the finalization of this work \cite{arxivversion}%
, we notice a recent paper studying non-Hermitian Aubry-Andr\'{e}-Harper
models \cite{ZengTopological2020}. We would like to point out that the
topological properties of our model originate only from the non-Hermitian
effects and thus require $\lambda \neq 0$, yet their model is topological
even when the non-Hermitian effects vanish $\lambda =0$. Our work reveals
that even pure non-Hermiticity in 1D could generate nontrivial topological
phases in 2D.

\begin{acknowledgments}
This work is supported by Air Force Office of Scientific Research
(FA9550-16-1-0387), National Science Foundation (PHY-1505496, PHY-1806227),
and Army Research Office (W911NF-17-1-0128). This work is also supported in
part by NSFC under the grant No. 11504285 and the Scientific Research
Program Funded by Natural Science Basic Research Plan in Shaanxi Province of
China (Program No. 2018JQ1058).
\end{acknowledgments}

\appendix

\section{Proof of Berry phase periodicity/quantization and the 2$\mathbb{Z}$
Chern number}

\label{AppA} In the main text, a few statements are made based upon physical
considerations and are supported with numerics. Here, more rigorous proofs
are provided for some important statements. In this section, we denote $%
k_{y}\equiv \phi _{L}$ to write the momentum-parameter space $(k,\phi _{L})$
as $\mathbf{k}=(k_{x},k_{y})$ for easy notation.

The right and left eigenvectors are defined as 
\begin{eqnarray}  \label{lrdef}
H(\mathbf{k})|\psi (\mathbf{k})\rangle _{R} &=&\omega (\mathbf{k})|\psi (%
\mathbf{k})\rangle _{R}, \\
H(\mathbf{k})^{\dagger}|\psi (\mathbf{k})\rangle _{L} &=&\omega ^{\ast }(%
\mathbf{k})|\psi (\mathbf{k})\rangle _{L},  \notag
\end{eqnarray}
with the normalization condition $_{L}\langle \psi (\mathbf{k})|\psi (%
\mathbf{k})\rangle _{R}=1$. There are four different definitions of Berry
connection $\mathcal{A}_{\mathbf{k}}^{ab}=-i_{a}\langle \psi|\partial _{%
\mathbf{k}}\psi\rangle _{b}$, where $a,b=L,R$. A natural generalization from
Hermitian systems would be $\mathcal{A}_{\mathbf{k}}^{LR}$, which is not
generally correct. While $\mathcal{A}_{\mathbf{k}}^{aa}$ is always real
because $_{a}\langle \psi|\partial_{\mathbf{k}}\psi\rangle _{a}$ is purely
imaginary due to the normalization condition, $\mathcal{A}_{\mathbf{k}%
}^{ab},a\neq b$ could be any complex number since the normalization
condition only restricts $_{a}\langle \psi|\partial _{\mathbf{k}%
}\psi\rangle_{b}+{}_{a}\langle \partial _{\mathbf{k}}\psi |\psi \rangle
_{b}=0$. To resolve this, we consider a normalized Berry connection 
\begin{equation}
\mathcal{A}_{\mathbf{k}}^{n}=\frac{1}{2}\left( \mathcal{A}_{\mathbf{k}}^{RL}+%
\mathcal{A}_{\mathbf{k}}^{LR}\right) ,
\end{equation}%
which is purely real because $\mathcal{A}_{\mathbf{k}}^{n}-\mathcal{A}_{%
\mathbf{k}}^{n\ast } =-\frac{i}{2}\big({}_{R}\langle \psi |\partial _{%
\mathbf{k}}\psi \rangle_{L}+{}_{L}\langle \psi |\partial _{\mathbf{k}}\psi
\rangle_{R}+{}_{L}\langle \partial _{\mathbf{k}}\psi |\psi
\rangle_{R}+{}_{R}\langle \partial _{\mathbf{k}}\psi |\psi \rangle _{L}\big)%
=-\frac{i}{2}\partial _{\mathbf{k}}\big({}_{R}\langle \psi |\psi
\rangle_{L}+{}_{L}\langle \psi |\psi \rangle _{R}\big)=0$ or simply $%
\mathcal{A}_{\mathbf{k}}^{LR}=\left(\mathcal{A}_{\mathbf{k}}^{RL}\right)^*$.
Such a well-defined Berry connection yields a (normalized) Berry phase $%
\gamma _{n}=\frac{1}{\pi}\oint d\mathbf{k\cdot }\mathcal{A}_{\mathbf{k}}^{n}$%
, which is real and quantized along a closed loop for any gapped
non-Hermitian systems.

We now show that such a Berry phase $\gamma _{n}(k_{y})=\frac{1}{\pi }\oint
dk_{x}\mathcal{A}_{k_{x}}^{n}(k_{y})$ would have a period $T$ if $%
H(k_{x},k_{y})=H^{\dagger }(k_{x},k_{y}+T)$ and the bands are real and
gapped (a weaker condition is that the bands are separable by their real
parts). Starting with the definition of left eigenvector in Equ.~\ref{lrdef}
and applying the aforementioned conditions, we have $H(k_{x},k_{y})|\psi
(k_{x},k_{y}+T)\rangle _{L}=\omega ^{\ast }(k_{x},k_{y}+T)|\psi
(k_{x},k_{y}+T)\rangle _{L}$ so that $\mathcal{A}_{k_{x}}^{RL}(k_{y})=%
\mathcal{A}_{k_{x}}^{LR}(k_{y}+T)$ and vice versa. Now, we realize that $%
\mathcal{A}_{k_{x}}^{n}(k_{y})=\frac{1}{2}\left( \mathcal{A}%
_{k_{x}}^{RL}(k_{y})+\mathcal{A}_{k_{x}}^{LR}(k_{y})\right) =\frac{1}{2}%
\left( \mathcal{A}_{k_{x}}^{LR}(k_{y}+T)+\mathcal{A}_{k_{x}}^{RL}(k_{y}+T)%
\right) =\mathcal{A}_{k_{x}}^{n}(k_{y}+T)$. Integrating along $k_{x}$, we
find the Berry phase has a similar periodicity $\gamma _{n}(k_{y})=\gamma
_{n}(k_{y}+T)$. Another consequence here is that the Chern number will be
the multiple of some integers due to the periodicity of Berry connection. We
could also define a normalized Chern number $C^{n}$ based on the normalized
Berry connection so that

\begin{eqnarray}
C^{n} &=&\frac{1}{2\pi}\oint dk_{x}dk_{y}\big(\partial _{k_{x}}\mathcal{A}%
_{k_{y}}^{n}-\partial _{k_{y}}\mathcal{A}_{k_{x}}^{n}\big) \\
&=&\frac{1}{2}\bigg(\frac{1}{2\pi}\oint dk_{x}dk_{y}\big(\partial _{k_{x}}%
\mathcal{A}_{k_{y}}^{RL}-\partial _{k_{y}}\mathcal{A}_{k_{x}}^{RL}\big) 
\notag \\
& &+\frac{1}{2\pi}\oint dk_{x}dk_{y}\big(\partial _{k_{x}}\mathcal{A}%
_{k_{y}}^{LR}-\partial _{k_{y}}\mathcal{A}_{k_{x}}^{LR}\big)\bigg)  \notag \\
&=&\frac{1}{2}\big(C^{RL}+C^{LR}\big)  \notag \\
&=&C^{ab}.  \notag
\end{eqnarray}

As $C_{T}^{n}=\frac{1}{2\pi}\oint dk_{x}\int_{k_{y}}^{k_{y}+T}dk_{y}\big(%
\partial _{k_{x}}\mathcal{A}_{k_{y}}^{n}-\partial _{k_{y}}\mathcal{A}%
_{k_{x}}^{n}\big)$ must be quantized due to the quantized charge pumping
along $k_{y}$ demanded by the periodicity of the Berry phase, it's easy to
see $C^{n}=N_{y}C_{T}^{n}$ belongs to a $N_{y}\mathbb{Z}$ class if $k_{y}$
has the period $k_{y}=k_{y}+N_{y}T$.

We also claim that the Berry phase is quantized at inversion-symmetric
points. To show this, we consider the symmetry $\mathcal{I}_{k}H(k_x,k_y)%
\mathcal{I}_{k}^{-1}=H(-k_x,-k_y+T)$. Such a symmetry dictates $H(k_x,k_y)|%
\mathcal{I}_{k}\psi(-k_x,-k_y+T)\rangle_R =\omega(-k_x,-k_y+T)|\mathcal{I}%
_{k}\psi(-k_x,-k_y+T)\rangle_R$. When the system is gapped, we have $%
\mathcal{A}^n_{k_x}(k_y)=\mathcal{A}^n_{-k_x}(-k_y+T)$, where the condition $%
\mathcal{I}_{k}^2=I$ is used. Then the Berry phase reads $\gamma_n(k_y)=%
\frac{1}{\pi}\oint dk_x\mathcal{A}^n_{k_x}(k_y)=\frac{1}{\pi}\oint dk_x%
\mathcal{A}^n_{-k_x}(-k_y+T)=-\frac{1}{\pi}\oint dk_x\mathcal{A}%
^n_{k_x}(-k_y+T)=-\gamma_n(-k_y+T)$. This implies that $\gamma_n(k_y^{\prime
})=-\gamma_n(k_y^{\prime })$ if $k_y^{\prime }=k_y^{\prime }-T$ (i.e., the
high-symmetry point), which means it can only take the quantized value 0 or
1.

We now go back to the specific model we discussed in the main text, where $%
T=\pi $ and $k_{y}=\phi _{L}$ has a period $2\pi $ ($N_{y}=2$). This means
that the Berry phase satisfies $\gamma _{n}(\phi _{L})=\gamma _{n}(\phi
_{L}\pm \pi )$ and the Chern insulator has a Chern number $2\mathbb{Z}$
which is determined by the symmetry $H(k,\phi _{L})=H^{\dagger }(k,\phi
_{L}+\pi )$. Another constraint on Berry phase is $\gamma _{n}(\phi
_{L})=-\gamma _{n}(-\phi _{L}\pm \pi )$ so that it is quantized at
inversion-symmetric point $\phi _{L}=\pm \frac{\pi }{2}$. Combing the two
conditions together, we find further $\gamma _{n}(\phi _{L})=\gamma
_{n}(\phi _{L}+\pi )=-\gamma _{n}(-\phi _{L})$ and thus, the Berry phase is
also quantized at $\phi _{L}=0$ and $\pi $. Note that the last relation is
imposed by the $\mathcal{Q}$ symmetry, which can be similarly proved if the
bulk band is gapped in the real part.

\section{Lifting band degeneracies in topological semimetal}

\label{AppB} In the topological semimetal phase, there could be band
degeneracy in momentum space, which hinders the computation of Berry phase
or $\mathbb{Z}_{2}$ invariant. In the following, we take the example $\alpha
=1/4$ for illustration purpose. The Bloch Hamiltonian reads 
\begin{eqnarray}
H &=&\frac{1+\cos (4k)+i\lambda \sin (4k)\cos (\phi _{L})}{2}\sigma
_{x}\otimes \sigma _{x} \\
&&+\frac{1-\cos (4k)-i\lambda \sin (4k)\cos (\phi _{L})}{2}\sigma
_{y}\otimes \sigma _{y}  \notag \\
&&+\frac{\sin (4k)+i\lambda \cos (\phi _{L})(1-\cos (4k))}{2}\sigma
_{x}\otimes \sigma _{y}  \notag \\
&&+\frac{\sin (4k)-i\lambda \cos (\phi _{L})(1+\cos (4k))}{2}\sigma
_{y}\otimes \sigma _{x}.  \notag
\end{eqnarray}%
The bulk bands are 
\begin{equation}
E_{\pm ,\pm }=\pm \sqrt{2-\lambda ^{2}\pm \sqrt{\frac{1}{2}(8-8\lambda
^{2}+\lambda ^{4}(1-\cos (4\phi _{L})))}}
\end{equation}%
at the high-symmetry point $k=0$. We notice that when $\cos (4\phi _{L})=-1$%
, i.e., $\phi _{L}=\frac{1}{4}\pi ,\frac{3}{4}\pi ,\frac{5}{4}\pi ,\frac{7}{4%
}\pi $, the band gap closes in momentum space with $E_{+,-}=E_{-,+}=0$ and $%
E_{+,+}=-E_{-,-}=\sqrt{2(2-\lambda ^{2})}$ when $\lambda <\sqrt{2}$,
manifesting the topological phase transition that was discussed in the main
text. There is always a degenerate point in the top $E_{+,\pm }=+\sqrt{%
2-\lambda ^{2}}$ and bottom $E_{-,\pm }=-\sqrt{2-\lambda ^{2}}$ two bands at 
$k=\frac{\pi }{4}$. Such a degeneracy prevents us from computing the
topological invariant in momentum space, while it is irrelevant to the
topological properties since it does not depend on $\phi _{L}$. To resolve
this, we apply a perturbation term on the asymmetric hopping 
\begin{equation}
t_{i+1,i}=1-\lambda \cos (2\pi \alpha i+\phi _{L}+\delta \phi _{L}),\delta
\phi _{L}\in \mathbb{R}
\end{equation}%
which perseveres the chiral symmetry so that it would not change the
associated topological invariant. It breaks the degeneracy at $k=\pi /4$ but
preserves the gap closings at topological phase transition points.

\begin{figure}[h]
\centering
\includegraphics[width=0.48\textwidth]{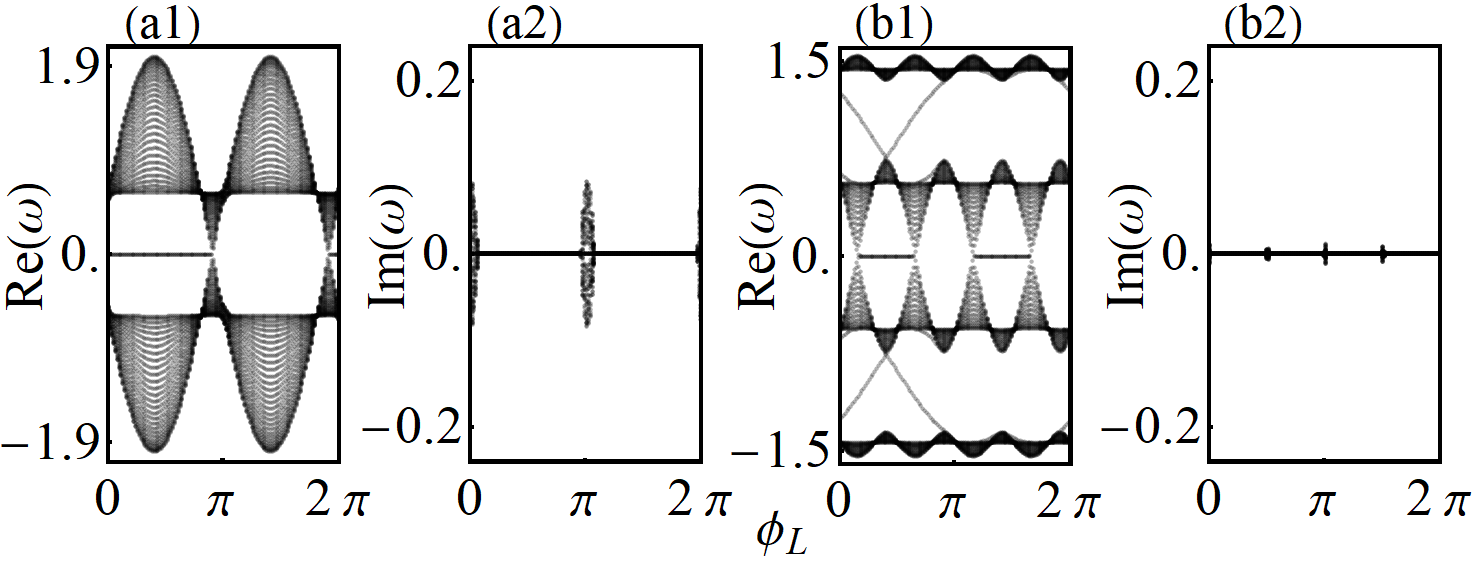}
\caption{(a) Open-boundary spectrum for varying $\protect\phi _{L}$ with $%
\protect\alpha =1/2$, $\protect\lambda =1$ and $\protect\delta \protect\phi %
_{L}=0.2\protect\pi $. (b) Similar as panel (a) but plotted with $\protect%
\alpha =1/4$.}
\label{SFigDeltaPhi}
\end{figure}

When a finite $\delta \phi $ is considered, a few interesting points raise.
First, the topological semimetal phase can be realized in a 2-level system,
i.e., $\alpha =1/2$ [see Fig.~\ref{SFigDeltaPhi}(a)], which represents some
non-Hermitian SSH models that have been discussed in previous literatures 
\cite{LieuS2018,YaoS2018}. The $\mathbb{Z}_{2}$ index reduces to the 1D Zak
phase of the ground state, which can correctly characterize the
non-Hermitian SSH model with chiral symmetry.

A finite $\delta \phi $ would spoil the 2$\mathbb{Z}$ Chern number and thus,
renders a $\mathbb{Z}$-type Chern insulator. Moreover, since it breaks the
degeneracy within the bottom and top two bands at $\alpha =1/4$
respectively, there would be a gap opening, providing the possibility of
supporting a nontrivial Chern insulator phase. We demonstrate these two
points in Fig.~\ref{SFigDeltaPhi}(b), where we do observe one chiral surface
wave (for a given propagating direction) residing within each gap. The zero
modes naturally persist but are shifted with the topological phase
transition points. The characterization of the topological properties are
similar and a case study of a mix phase is presented in Appendix~\ref{AppD}.

\begin{figure}[h]
\centering
\includegraphics[width=0.48\textwidth]{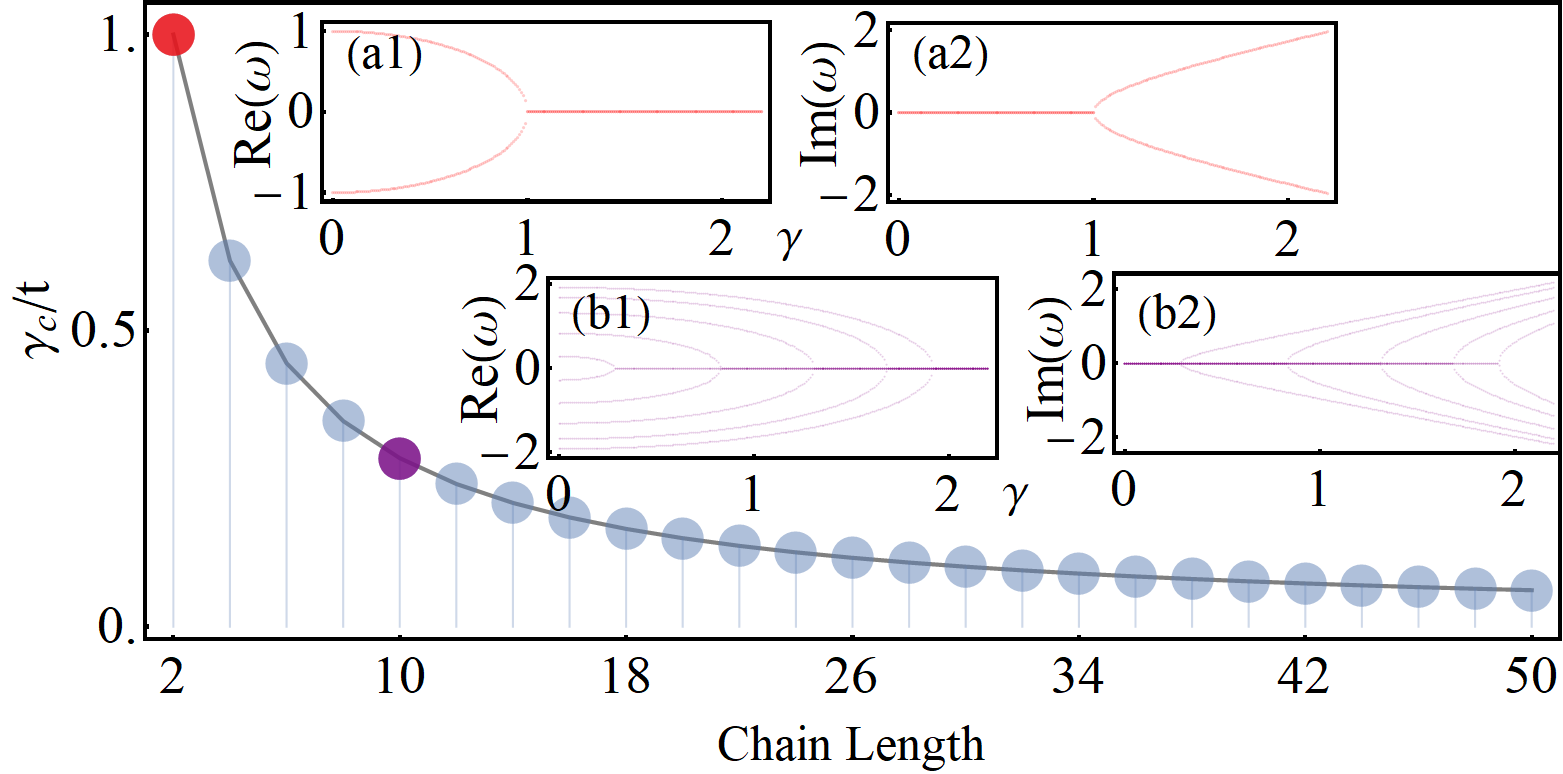}
\caption{Critical gain/loss rate for $\mathcal{PT}$ breaking versus the
length of the total chain in real space. The insets show how the spectra
change with $\protect\gamma $ at given chain lengths 2 and 10.}
\label{SFigPTBreaking}
\end{figure}

\section{$\mathcal{PT}$-symmetric phase by gain/loss is fragile}

\label{AppC} In main text, we state that the usual $\mathcal{PT}$-symmetric
phase by gain/loss is fragile, therefore the spectrum cannot be purely real
in the separable regime \cite{YuceC2014}. To illustrate the fragile $%
\mathcal{PT}$-symmetric phase, we consider a simple model described by the
following tight-binding Hamiltonian 
\begin{equation}
H_{PT}=t\sum_{i}\left( \hat{c}_{i}^{\dagger }\hat{c}_{i+1}+h.c\right)
+(-1)^{i}\gamma \hat{c}_{i}^{\dagger }\hat{c}_{i},
\end{equation}%
where $\gamma >0$ is on-site gain/loss rate. The Hamiltonian is defective at
the exceptional point and the system enters the partially broken regime
(part of the spectrum becomes complex) at the smallest $\gamma _{c}$
satisfying $\text{Det}(H_{PT}(\gamma _{c}))=0$. The determinant can be
solved through the recursive equation $D_{n}=(-1)^{n}\gamma
D_{n-1}-t^{2}D_{n-2}$ with boundary conditions $D_{1}=-i\gamma $ and $%
D_{2}=\gamma ^{2}-t^{2}$. The critical values $\gamma _{c}$ for different
chain lengths are shown in Fig.~\ref{SFigPTBreaking}, where we see a fast
drop of $\gamma _{c}$ when the chain length starts to increase and
ultimately approaches zero. In the insets (a) and (b), we show the spectrum
for varying $\gamma $ at different chain lengths 2 and 10. This argument
also applies to more general cases such as larger unit cell, quasiperiodic
potential, or higher dimensions. This explains the complex edge states
observed in many models with on-site gain and loss \cite%
{YuceC2014,TakataK2018}.

Such a fragility does not happen when the non-Hermiticity is introduced by
asymmetric hopping as illustrated in Fig.~\ref{FigCoExist}(a) and (b), thus
our model enjoys real edge modes and is more accessible to experimental
observation.

\begin{figure}[h]
\centering
\includegraphics[width=0.48\textwidth]{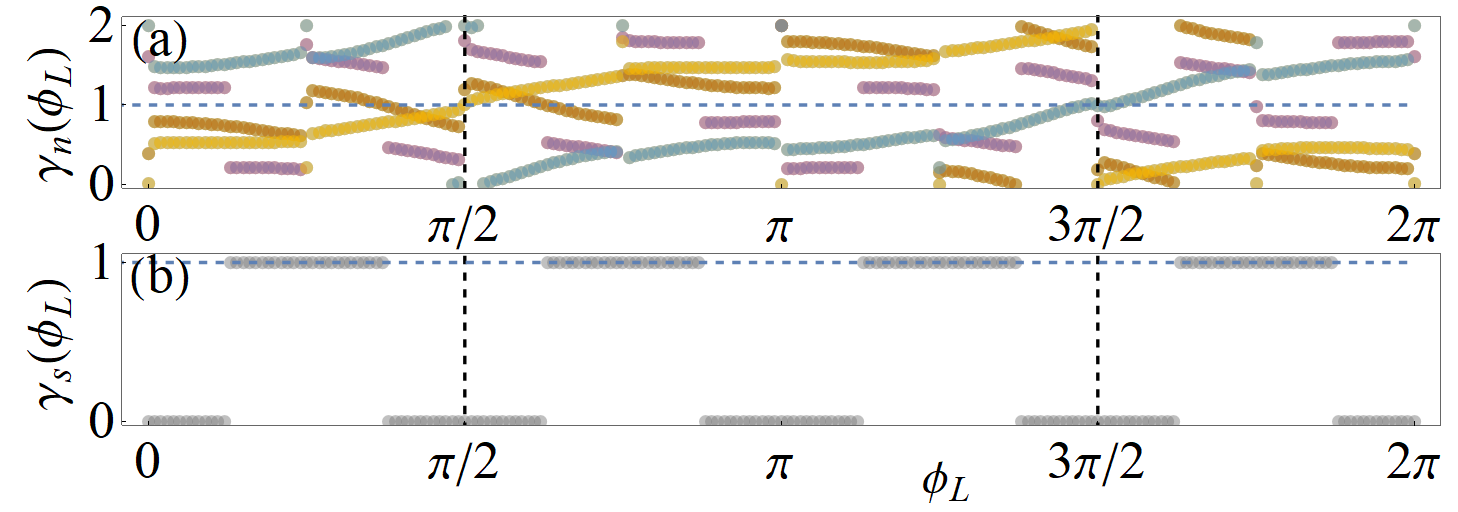}
\caption{(a) Berry phase $\protect\gamma _{n}(\protect\phi _{L})$ for the
coexisting phase shown in Fig.~\protect\ref{FigCoExist}(e). The dashed
vertical lines indicate inversion-symmetric points and the numeric errors
are more significant near gap closing points. (b) The $\mathbb{Z}_{2}$
invariant computed from panel (a).}
\label{SFigCoExist}
\end{figure}

\section{Topological characterization of the coexisting phase}

\label{AppD} Although we have studied two topological phases individually,
the coexisting phase can be characterized in a similar way if the band
degeneracy is treated carefully. We still consider a small perturbation $%
\delta \phi _{L}$ when compute the Berry phase and the results are shown in
Fig.~\ref{SFigCoExist}(a). Due to the gapless phase, the Berry phase does
not have the periodicity any more. However, it is still quantized at
inversion-symmetric point because a small $\delta \phi _{L}$ only perturbs
the system slightly. In Fig.~\ref{FigCoExist}(e), the chiral surface wave
still crosses at the high-symmetry points, at which the Berry phases are
quantized to non-trivial values [Fig.~\ref{SFigCoExist}(a)].

We notice that the Chern number is not well-defined due to the topological
phase transition point and degeneracies between top/bottom four bands. While 
$\delta \phi _{L}$ term allows computing the Chern number, it breaks the 2$%
\mathbb{Z}$ constraint, and the Chern numbers are odd for the top/bottom two
bands (see also the charge pumping in Fig.~\ref{SFigCoExist}(a)). When $%
\delta \phi _{L}$ is gradually tuned to 0, the band Chern number should not
change because there is no topological phase transition. When $\delta \phi
_{L}=0$, two bands become degenerate. If we take both bands as a single
band, the Chern number is even and the chiral surface wave appears in pair
in each gap.

In comparison, the zero modes can be directly characterized by the $\mathbb{Z%
}_{2}$ invariant, which is shown in Fig.~\ref{SFigCoExist}(b), and it is
consistent with the zero modes observed in Fig.~\ref{FigCoExist}(e).

\end{document}